\documentclass[5p]{elsarticle}

\usepackage{graphics}
\usepackage{eufrak}
\usepackage{amsmath,amssymb}
\usepackage{amsmath,color}
\usepackage{graphicx, hyperref}
\usepackage{psfrag}
\usepackage{textcomp}
\usepackage[usenames,dvipsnames]{xcolor}

\biboptions{sort&compress}

\newcommand{\be}{\begin{equation} }
\newcommand{\ee}[1]{ \label{#1} \end{equation}}
\newcommand{\ba}{\begin{eqnarray} }

\newcommand{\ea}[1]{ \label{#1} \end{eqnarray}}

\newcommand{\pd}[2]{ \frac{\partial #1}{\partial #2} }
\newcommand{\ka}{\zeta}
\newcommand{\prd}{Phys.~Rev.~D}

\newcommand{\prl}{Phys.~Rev.~Lett.~}

\begin{document}

\title{Black hole horizons can hide positive heat capacity}

\author[wigner]{Tam\'as S.~Bir\'o}
\ead{biro.tamas@wigner.mta.hu}
\author[wigner,ist]{Viktor G.~Czinner}
\ead{czinner.viktor@wigner.mta.hu}
\author[nihon]{Hideo Iguchi}
\ead{iguchi.h@phys.ge.cst.nihon-u.ac.jp}
\author[wigner,bme]{P\'eter V\'an}
\ead{van.peter@wigner.mta.hu}

\address[wigner]{HAS Wigner Research Centre for Physics, H-1525 Budapest, P.O.Box 49, Hungary}
\address[ist]{Centro de Astrof\'{\i}sica e Gravita\c c\~ao, Departamento de F\'{\i}sica, 
Instituto Superior T\'ecnico,\\ Universidade de Lisboa, Av.~Rovisco Pais 1, 1049-001 Lisboa, Portugal}
\address[nihon]{Laboratory of Physics, College of Science and Technology, Nihon University, 
274-8501 Narashinodai, Funabashi, Chiba, Japan}
\address[bme]{Department of Energy Engineering, Budapest University of Technology and
Economics, Bertalan Lajos u.~4-6, 1111 Budapest, Hungary}

\date{\today}

\begin{abstract}
Regarding the volume as independent thermodynamic variable we point out that black hole 
horizons can hide positive heat capacity and specific heat. Such horizons are mechanically 
marginal, but thermally stable. In the absence of a canonical volume definition, we consider 
various suggestions scaling differently with the horizon radius. Assuming Euler-homogeneity 
of the entropy, besides the Hawking temperature, a pressure and a corresponding work term 
render the equation of state at the horizon thermally stable for any meaningful volume concept 
that scales larger than the horizon area. When considering also a Stefan--Boltzmann radiation 
like equation of state at the horizon, only one possible solution emerges: the Christodoulou--Rovelli 
volume, scaling as $V\sim R^5$, with an entropy $S = \frac{8}{3}S_{BH}$.
\end{abstract}


\maketitle

\section{Introduction}

The irreducible mass of black holes is connected to an entropy function in black hole 
thermodynamics \cite{Bek72t,Bek72a,Bek73a,Bek74a,BarEta73a}. This relation inspired many 
further investigations about the origin of the fundamental equations including various 
ideas toward quantum gravity 
\cite{Jac95a,StroVafa,LQGent1,LQGent2,Pad10a,Pad14a,Ver11a,Ver17a,BirVan15a}. 
It is well known that the related equation of state has some peculiar properties from a 
thermodynamic point of view. Due to the fact that the irreducible mass of black holes is 
proportional to the radius of their event horizon, the entropy, proportional to its surface, 
$S(M) \sim M^2$, is seemingly convex and the heat capacity derived from it is negative. 
This is common in all bound systems where the total energy is negative and the
kinetic energy is positive, then due to an increase of the temperature via an increase in
the kinetic energy, -- in a stationary state satisfying a virial theorem, -- the total energy will
decrease, displaying formally a negative heat capacity. A thermal equilibrium between a negative
specific heat system and a positive one  is, however, not possible.
Black holes in this sense seem thermally unstable.

There are various suggestions that could counterbalance 
the consequent mechanical instability \cite{Thi70a,Haw74a,Dav78a,BhaSha17a}, however, 
its very existence is an obstacle in constructing reasonable statistical theories for black holes 
\cite{LavDun90a,Sre93a,tHo85a,ZurTho85a}. 
A careful distinction of extensivity and additivity in the related thermostatistics promises to give an 
insight into the problem \cite{Azr14a,TsaCir13a}, and a R\'enyi entropy \cite{Ren61a}  based theory 
actually removes the convexity of the Bekenstein--Hawking entropy of black holes \cite{BirCzinIg,CziIgu16a,CziIgu17}. 

In this Letter we demonstrate that the black hole horizon entropy formula is concave if treated as 
a function of at least two variables, and leads to ''normal'' thermodynamic behavior, with positive 
specific heat and marginal mechanical stability. We argue that considering any reasonable volume 
concepts (e.g.~the Parikh \cite{Par06a} or the Christodoulou--Rovelli definition \cite{ChrRov15a})
as an independent thermodynamical variable together with the related homogeneity assumption, eliminates 
the inconsistency while keeping the original formula. 


First a brief review of the custom derivation is given which leads to the {currently accepted}
conclusion of assigning negative heat capacity to such objects. Then we derive the thermodynamic 
properties of Schwarzschild black holes by including the usual work term in the first law based only 
on the assumption that the entropy is a first order homogeneous (extensive) function of the volume. 
Throughout this work we use units such as $\hbar=G=c=k_B=1$.

\section{Black hole EoS with volume term}

The traditional presentation of the negative heat capacity problem is as follows: 
Schwarzschild black hole horizons have a radius of $R=2M$, and a Bekenstein--Hawking entropy
of a quarter of the horizon area
\be
 S \: = \: {\pi R^2}.
\ee{BHENTROP}
Since the internal energy is dominated by the mass energy producing the same horizon, $E=M=R/2$, 
one light-heartedly considers a curious equation of state:
\be
 S(E) \: = \: {4\pi} \, E^2.
\ee{WRONGEOS}
This ``equation of state'' has strange properties. The absolute temperature, determined from
\be
 \frac{1}{T} \: = \: \frac{dS}{dE} \: = \: 8\pi \, E,
\ee{WRONGTEMP}
is growing with decreasing energy. This discrepancy {results} a negative heat capacity 
signalling {\em thermal instability} in the traditional view:
\be
 - \frac{1}{CT^2} \: = \: \frac{d^2S}{dE^2} \: = \: 8\pi \: > \: 0,
\ee{WRONGC}
which leads to the conclusion of having $C=-2S< 0$.
Negative  heat capacity occurs in all systems having negative total energy.
It is questionable, however, whether the total energy has to be counted as
internal energy when deriving thermal properties of a system.

Here we present an alternative approach which is thermodynamically consistent, and free from 
such oddities. First of all we consider the volume, enclosed by the event horizon, as a further 
thermodynamical variable. The physical volume of a black hole has been a long standing problem 
in general relativity. The standard definition operates with surfaces of simultaneity and therefore 
it is a strongly coordinate dependent notion. Recently, Christodoulou and Rovelli introduced an 
elegant, geometric invariant definition \cite{ChrRov15a}, where the volume of a Schwarzschild black 
hole has been defined as the largest, spherically symmetric, spacelike hypersurface $\Sigma$ bounded 
by the horizon. The corresponding CR-volume (when the thermal property of the Hawking radiation 
\cite{Hawking1974sw} is also taken into account) scales as $V\sim R^5$, which for an astrophysical 
black hole turns out to be very large indeed. This result motivated further investigations about the 
role this volume may play in the thermodynamic behavior of black holes \cite{ChrDeL16a,Bengtsson,Ong1,Ong2,Zha17a},
in particular, based only on simple causality considerations, Rovelli argues \cite{Rovelli2017} that black holes 
should have more states than those giving the Bekenstein--Hawking entropy, and the CR-volume is 
large enough to store these entropic states.

In this Letter we consider the phenomenological consequences of the volume scaling of the black hole
entropy, however we do not restrict our investigations to the CR-measure only. The approach taken 
here is completely general and valid for any meaningful volume definition. We will show, however, 
that by considering a Stefan--Boltzmann radiation like equation of state at the horizon (arising 
naturally from a Hawking radiation), the CR-volume scaling is reproduced.

The step to consider the volume as a thermodynamic variable is a fundamental one which also associates 
a pressure to the event horizon. In standard thermodynamics there exist a relationship (the Gibbs--Duhem 
relation (see e.g.~\cite{Callen})) among the intensive parameters of a system which is a consequence of 
the first order homogeneous property of the entropy function. This homogeneity relation is not valid within 
the standard picture of black hole mechanics (see e.g.~\cite{TsaCir13a,Bravetal} and references therein). 
A modification by York and Martinez \cite{Yor86a,Mar96a1,Mar96a2} tries to separate the surface of the 
horizon as an independent thermodynamic variable, however, the consequent scaling relations are not first 
order Euler-homogeneous, therefore there is no real Gibbs--Duhem relation in that framework \cite{Wri80a}. 

In the present approach, separating the volume to be the independent thermodynamic variable naturally resolves
the Gibbs--Duhem relation issue, which, together with the well-known power-law scaling of the energy, $E$, the total 
entropy, $S$, and the volume, $V$, with the horizon radius, $R$ of a Schwarzschild black hole, naturally 
suggests the general class of equation of states in the form  
\be
 S(E,V) \: = \: \zeta \, E^{\alpha} \, V^{\beta},
\ee{BASICEOS}
and the Euler-homogeneity assumption sets the condition $\alpha + \beta = 1$. 
 This form of equation of state does not contradict to the ''no hair'' theorem \cite{Israel},
as long as both $E(M)$ and $V(M)$ depend only on the sole physically relevant property
of a Schwarzschild black hole, its mass $M$. Nevertheless $S(E,V)$ has to be handled as a two-variable
function when obtaining its partial derivatives, and their corresponding physical interpretation.
Only these have to be taken at the end on physical line discribed by the pair $(E(M),V(M))$
in the paremeter space.
Temperature, partial derivative against $E$, is no more or less physical then pressure, obtained
from partial derivative against $V$. Microscopically both the absolute temperature and pressure
are positive in kinetic theories, while the classical pressure may turn out to be negative
in bound systems. In those cases the quantum uncertainty may stabilize such systems.
But this very same actor is responsible for the Unruh-type Hawking temperature.

In order to further specify the black hole equation of state by keeping the power-law form and without the loss 
of generality, one can parametrize the volume as
\be   V \: = \: R^{c+3} \, I_V, \ee{SCALING}
where $I_V$ is constant, independent of the horizon radius. For any choice of $c \ne 0$ the volume  in the present 
context is not the Euclidean three-volume, usually considered in everyday thermodynamics. According to the 
Schwarzschild black hole picture, the required dependence of the total energy on the radius, $E =M = R/2$, and 
the total entropy is proportional to the horizon area,  $S = 4\pi \lambda R^2 =\pi \lambda M^2$, 
where $\lambda = 1/4$ for the Bekenstein--Hawking entropy.

The scaling of the volume with the radius, i.e.~the parameter $c$, remains undetermined so far. 
The parameter $\lambda$ together with $I_V$ stays also undetermined at this level. From the equation 
of state (\ref{BASICEOS}) we have
\be4\pi \lambda R^2 \: = \: \zeta \, (R/2)^{\alpha} \, (I_VR^{c+3})^{\beta} \ee{BASICPOWER}
and therefore
\be
 2 \: = \: \alpha + \beta(c+3).
\ee{BASICPOWERS}
For further specification of the parameters we need more input from the physical picture.
Calculating the thermodynamical derivatives of $S(E,V)$ one interprets the temperature
\be
 \frac{1}{T} \: = \: \pd{S}{E} \: = \: \alpha \zeta E^{\alpha-1}V^{\beta} \: = \: \alpha \frac{S}{E} \: 
 = \: 8\pi \lambda \alpha R.
\ee{ONEPERTEMP}
This temperature $T$ is to be equal to the Hawking temperature \cite{Hawking1974sw}, $T_H=1/(4\pi R)$, 
which is the Unruh temperature \cite{UCP}, belonging to the gravitational acceleration at the horizon 
(without the red-shift factor). Keeping this equality delivers $\lambda=1/(2\alpha) $.
The other partial derivative,
\be
 \frac{p}{T} \: = \: \pd{S}{V} \: = \: \beta \zeta E^{\alpha}V^{\beta-1} \: = \: \beta \frac{S}{V},
\ee{PRESPERTEMP}
leads to another form of the equation of state, that is generally more useful in hydrodynamical calculations,
\be
 p \: = \: \frac{\beta}{\alpha} \, \frac{E}{V}.
\ee{PEOS}
The classical choice of $\beta=0$ in (\ref{BASICEOS}) leads to zero pressure, $p=0$. However, as it has been 
demonstrated by various authors \cite{UCP,Candelas,Page}, a nonvanishing pressure at the event horizon is always 
expected originating e.g.~from vacuum polarization effects in semi-classical approximations to Einstein's theory.
Furthermore, the Hawking radiation \cite{Hawking1974sw} also implies a Stefan--Boltzmann radiation-like equation 
of state at the horizon with nonzero pressure.

\section{Specific heat and stability}

In order to show that black holes can have a positive sepcific heat, we consider the second partial derivative 
of the entropy against the energy. The definition:
\be
 \pd{^2S}{E^2} \: = \: \pd{}{E} \frac{1}{T} \: = \: - \frac{1}{Vc_VT^2}
\ee{CVFROMSEE}
compared with (\ref{BASICEOS}) results in
\be
 \pd{^2S}{E^2} \: = \: \alpha (\alpha-1) \frac{S}{E^2}.
\ee{BASICSEE}
From this comparison, using $1/T=\alpha S/E$,  the following solution emerges:
\be
 c_V \: = \: \frac{\alpha}{1-\alpha} \, \frac{S}{V},
\ee{CVPOWER}
which can be positive when $0 < \alpha < 1$.
Comparing with the general form of the pressure with the power scaling ansatz, 
we obtain the relation:
\be
 c_V \cdot p \: = \: \frac{\beta}{(1-\alpha)} \, \frac{1}{V^2} \, E \cdot S.
\ee{NICERELATION}
Euler-homogeneity requires $\alpha+\beta=1$, which renders the ratio $\beta/(1-\alpha)$ 
also to be one. With positive entropy, positive energy and pressure, the specific heat at 
constant volume is {\em necessarily positive}. Taking into account (\ref{BASICPOWERS}) 
delivers
\be
 \alpha \: = \: \frac{c+1}{c+2}  { \qquad \mbox{and} \qquad \beta \: = \: \frac{1}{c+2}.}
\ee{EOSHOMO}
Therefore the specific heat in (\ref{NICERELATION}) is always positive when $c > -1$, i.e.~when
the volume scales with the horizon radius larger than the surface area.

In addition to Euler-homogeneity, by requiring the 3-dimensional radiation formula, one considers 
$\alpha = 3 \beta$. Together with (\ref{BASICPOWERS}) this results
\be
  \alpha \: = \: \frac{6}{6+c}  { \qquad \mbox{and} \qquad \beta \: = \: \frac{2}{6+c}.}
\ee{EOSHOLO}
The only solution which satisfies Euler-homogeneity (\ref{EOSHOMO}) and the radiation equation 
of state (\ref{EOSHOLO}) requirements at the same time is $c=2$, which results in a $V\sim R^5$ 
volume scaling, just like the Christodoulou--Rovelli volume \cite{ChrRov15a,Zha17a}. In this case 
$\alpha=3/4$, $\beta=1/4$ and the entropy of the black hole is still proportional to the horizon
area \hbox{$S\sim E^{3/4}V^{1/4} \sim R^{3/4} R^{5/4} \sim R^2$}, although the factor, $\lambda=2/3$ 
leads to a slightly larger coefficient than the one in the classical Bekenstein--Hawking formula. 
This result, however, has the clear advantage of having a positive specific heat.

In equating the expressions for $\alpha$ of (\ref{EOSHOMO}) and (\ref{EOSHOLO}) provides another 
possible solution, the $c=-3$. This results in a constant volume factor and leads to $\alpha=2$ 
for any $\beta$ from (\ref{BASICPOWERS}) independent of the conditions (\ref{EOSHOMO}) and 
(\ref{EOSHOLO}). $\alpha=2$ means also that $\lambda=1/4$, which reproduces the Bekenstein--Hawking 
formula $S=\pi R^2$. This choice, however, is not a real solution of the problem as it can never 
satisfy the conditions (\ref{EOSHOMO}) and (\ref{EOSHOLO}) for $\beta$ simultaneously. For example, 
it provides $\beta=2/3$ from the radiation equation of state (\ref{PEOS}), while $\beta=-1$ from Euler-homogeneity.
More importantly, as it is well known, this choice also leads to a negative specific heat.

\section{Causality and the third law of thermodynamics}

Based on this possibility of a thermodynamically stable scenario for black holes, it is intriguing to discuss 
certain aspects of it. Various scalings of the thermodynamically relevant volume with the horizon radius 
-- although cannot change our conclusion about a 
positive specific heat, formulated in (\ref{NICERELATION}) -- give us the possibility of different translations 
of the entropic equation of state, $S(E,V)$ to the more common mechanic equation of state, $p(E/V)$.

The most naive assumption (not solving our requirements though) deals with $c=0$. In  this case $V \sim R^3$, 
as this were the case in Euclidean geometry of the three-space. We note here however, that this scaling 
is also valid for a much wider class of geometries (see e.g. {\cite{Par06a}}). This choice would lead to
\be
 p \: = \: E/V \: = \: \epsilon \qquad {\rm and} \qquad c_V \: = \: S/V \: = \: \mathfrak{s}.
\ee{NAIVECHOICE}
While this scenario appears as thermally perfectly stable, it represents the allowed most extreme pressure without
violating causality, i.e.~it conjectures a velocity of sound equal to that of the light: $dp/d\epsilon=1$.
{We note here that any $c<0$ model, among others assuming a surface-shell as the relevant
volume with $c=-1$, would lead to an equation of state with an acausal speed of sound, $dp/d\epsilon > 1$.
From (\ref{PEOS}) $dp/d\epsilon=\beta/\alpha=1/(c+1)$, diverges for $c=-1$.

Finally, the temperature dependence of energy density and pressure with assumed Euler-homogeneity connects our 
result to more customary views. Expressing these quantities one obtains
\be
 \frac{E}{V} \, = \, \epsilon \: = \: \sigma_c \, T^{c+2}  \qquad  {\rm and} \qquad  p \: = \: \frac{1}{c+1} \sigma_c \, T^{c+2}.
\ee{TEMPEOS}
Here $\sigma_c=(\ka b/a)^{c+2}$ is the corresponding ``Stefan--Boltzmann constant'' for a far observer.
It is also worth noting that the specific heat, expressed with the temperature,
\be
 c_V \: = \: \ka \, \sigma_c^{b/a} \, (c+1) \, T^{c+1},
\ee{SPECHEATEMP}
reveals that the thermodynamical view presented here {\em also satisfies the third law}: at $T=0$, also
$c_V=0$ for any $c > -1$ choice.

Again, the naive volume scaling with $c=0$, however physically allowed, would lead to the strange conclusion 
$p=\epsilon \sim T^2$, $c_V=\mathfrak{s} \sim T$, but this is all physical and thermally stable. On the other 
hand, arguments assuming a traditional Stefan--Boltzmann radiation like equation of state (based on the 
thermal property of the Hawking radiation \cite{Hawking1974sw}) are built on $p = \epsilon/3 \sim T^4$. This 
immediately requires $c=2$, and leads to a volume measure scaling like $V \sim R^5$. Indeed, as shown above,
this power is in perfect agreement with the results of the Christodoulou-Rovelli 
volume \cite{ChrRov15a,ChrDeL16a} together with the black body spectrum of the Hawking radiation 
\cite{Hawking1974sw,Zha17a}.

According to the original idea of the Hawking radiation \cite{Hawking1974sw}, the scaling volume would be a surface, 
and hence one would consider $c=-1$. As seen before, the specific heat is negative in this case. For $c=-1+0^+$ 
our stability arguments nevertheless hold. The causality problem of sound waves, however, remains for all $c < 0$ 
models. 

\section{Conclusions}

Extensivity, rigorously distinguished from additivity \cite{Mar96a2,Tou02a,Tsa09b} is represented 
by first order Euler-homogeneity of the entropy by any of its state variables. This is necessary to 
introduce thermodynamic densities for fields \cite{BerVan17b}. Any meaningful concept of black hole 
volume requires reconsidering black hole thermodynamics, including the homogeneity relations as well.
We showed that standard thermodynamic properties, i.e.~homogeneity and volume scaling, 
are both compatible with the classic result that the black hole's entropy is proportional with 
the horizon area. Our approach naturally modifies the longstanding issues related to negative heat capacity 
and thermal instability, while the Hawking radiation formula also singles out the Christodoulou--Rovelli 
volume and the $\lambda=2/3$ coefficient factor as physical quantities from the free parameters of the theory.
As for the description of presenting an equation of state on the horizon while observing it only from a 
far distance, we are also in accord with phenomenological approaches to black hole thermodynamics. Based 
on this picture a Hawking pressure may well be associated to the Hawking temperature at black hole horizons.

Apart form the stability issue, there are several important problems where our extended thermodynamic 
background can also give a deeper insight. The connection to the cosmological term in the Einstein 
equation, for example, has already shown to be consistent with a thermodynamic interpretation using 
volume and pressure \cite{Azr15a}. The extension to AdS and more general spacetimes leads to 
further consequences \cite{Kastor09,HenEta17a}. The recently suggested complexity-volume relation 
demonstrates that holography can also be connected to volume changes \cite{BroEta16a,CouEta17a,HanEta17a}. 

Generalizations of this discussion for charged, rotating and even more general black holes shall be 
postponed to follow-up works. Based on some very recent, exciting experimental results \cite{ndim1,ndim2} 
on the possible existence of higher dimensions however, the following outlook may be 
instructive. By considering a $d$-dimensional radiation pressure, one would have $\beta/\alpha=1/d$, 
which would replace $c+3$ by $c+d$ in the above derivations. Satisfying Euler-homogeneity and having 
a power-like equation of state leads to $c=2$ and $c=-d$ as formal solutions, i.e.~to $V\sim R^{d+2}$
and $V\sim $ constant. This result distinguishes again the Christodoulou--Rovelli scenario for black holes 
in all spatial dimensions.

\section*{Acknowledgement}
This work was supported by the National Research, Development and Innovation Office – NKFIH 
under the grants 116197, 116375, 124366 and 123815. V.G.Cz.~thanks to Funda\c c\~ao para a Ci\^encia 
e Tecnologia (FCT) Portugal, for financial support through Grant No.~UID/FIS/00099/2013. The authors 
thank to L\'aszl\'o {B.}~Szabados and M\'aty\'as Vas\'uth for valuable discussions.  

\bibliographystyle{unsrt}

\end{document}